\documentclass[prb,twocolumn,showpacs,superscriptaddress,floatfix]{revtex4-2}
\usepackage{amssymb,amsfonts,amsmath}
\usepackage{adjustbox}
\usepackage{color}
\usepackage{xr}
\usepackage{braket}
\usepackage{xcolor}

\begin{document}
\title{A New Two-Dimensional Dirac Semimetal Based on the Alkaline Earth Metal, CaP$_3$}

\author{Seoung-Hun Kang}
\affiliation{Materials Science and Technology Division, Oak Ridge National Laboratory, Oak Ridge, TN 37831, USA}
\author{Wei Luo}
\affiliation{Center for Nanophase Materials Sciences, Oak Ridge National Laboratory, Oak Ridge, TN 37831, USA}
\author{Sinchul Yeom}
\affiliation{Materials Science and Technology Division, Oak Ridge National Laboratory, Oak Ridge, TN 37831, USA}
\author{Yaling Zheng}
\affiliation{Department of Physics and Astronomy, University of Tennessee, Knoxville, TN 37996, USA}
\author{Mina Yoon\thanks{corresponding author}}
\email{myoon@ornl.gov}
\affiliation{Materials Science and Technology Division, Oak Ridge National Laboratory, Oak Ridge, TN 37831, USA}

\begin{abstract}
Using an evolutionary algorithm in combination with first-principles density functional theory calculations, we identify two-dimensional (2D) CaP$_3$ monolayer as a new Dirac semimetal due to inversion and nonsymmorphic spatial symmetries of the structure. This new topological material, composed of light elements, exhibits high structural stability (higher than the phase known in the literature), which is confirmed by thermodynamic and kinetic stability analysis. Moreover, it satisfies the electron filling criteria, so that its Dirac state is located near the Fermi level. The existence of the Dirac state predicted by the theoretical symmetry analysis is also confirmed by first-principles electronic band structure calculations. We find that the energy position of the Dirac state can be tuned by strain, while the Dirac state is unstable against an external electric field since it breaks the spatial inversion symmetry. Our findings should be instrumental in the development of 2D Dirac fermions based on light elements for their application in nanoelectronic devices and topological electronics.
\end{abstract}


\maketitle

\section{Introduction}
Electronic band structures are fingerprints of the crystal symmetry of solids~\cite{fu2011a}, determining their band degeneracies and band topologies. The crystal symmetry becomes more prominent in accommodating Dirac state in low-dimensional system, such as two-dimensional (2D) atomic thick material like graphene~\cite{castro2009a}. As the Dirac state in the graphene is guaranteed by crystal symmetry, spin-orbit coupling (SOC) breaks the Dirac state. Resulting in a small band gap~\cite{haldane1998a,kane2005a}, thus graphene becomes a topological insulator~\cite{kane2005b}. Many other 2D materials such as silicene~\cite{cahangirov2009a,liu2011a}, germanene~\cite{cahangirov2009a,liu2015a}, 2D boron and carbon allotropes~\cite{malko2012a,xu2014a,zhou2014a,ma2016a,jiao2016a,feng2017a}, group-VA phosphorus structures~\cite{kim2015a,lu2016a,kang2019a}, and 5d transition metal trichoride~\cite{sheng2017a} have similar properties to graphene.

Theoretical studies have so far predicted many 2D topological states~\cite{kang2019a,Guan2017a,zhongfei2019a,deping2022a}. Among the states of fundamental interest are those protected by the interplay of different symmetries. For example, a 2D Dirac state is predicted to be guaranteed if the system with nonsymmorphic space group maintains both inversion and time-reversal symmetries (TRS)~\cite{young2015a}. Another interesting state is the 2D Dirac state for a system with two nonsymmorphic space group symmetries, where the Dirac state is at the time-reversal invariant momenta (TRIMs)~\cite{kang2019a}. One of the challenges in the field is to bring the Dirac states close to the Fermi level (E$_F$) so that the topologically protected state can be observed and used in experiments. In the pristine structure, the location of Dirac state is primarily determined by the electron filling~\cite{young2015a,wieder2016a}. Many of the 2D materials that stabilize the Dirac state do not satisfy the partial filling state of the electrons, so their energy levels are far from the E$_F$~\cite{lu2016a,kang2019a,singh2014a}.  While the position of the Fermi level can be controlled by mechanical strain~\cite{pereira2009a} or chemical doping~\cite{panchokarla2009a}, finding the desirable electronic properties of thermodynamically stable structures could be another challenge. 

In this paper, we report a new 2D Dirac semimetal as a result of inversion and nonsymmorphic spatial symmetries of the structure. It is a new monolayer structure of CaP$_3$ with the Dirac state near the Fermi level. The structure is discovered by performing first-principles calculations combined with an evolutionary algorithm. Theoretical symmetry analysis agrees well with the first-principles band structures and proves the existence of two Dirac states protected by inversion and nonsymmorphic space symmetries at TRIMs in the Brillouin zone (BZ) boundary. The new structure proves to be dynamically and thermodynamically stable, as indicated by our phonon analysis and molecular dynamics (MD) simulations at high temperature (500~K). In addition, we investigate the stability of the Dirac state by applying various symmetric lattice strains, such as biaxial, uniaxial, and shear strains, and confirm that the properties of the Dirac state are intact under these stains, but its location. Therefore, we propose strain as an effective method to tune the location of the Dirac state. Our result demonstrate how crystal symmetry plays a role in stabilizing Dirac states in 2D materials and the influence of external parameters to tune the electronic properties to the desirable outcome.

\section{Calculational approaches}
To identify thermodynamically stable structures of 2D CaP$_3$, we perform crystal prediction calculations using the particle swarm optimization (PSO) algorithm implemented in Crystal Structure Analysis by Particle Swarm Optimization (CALYPSO)~\cite{Yachao2010a, luo2021a}. In the beginning of the simulation, we start with 20 random structures within 2D space groups, and the structural evolution proceeds up to 6 generations based on the PSO scheme. At the end of the simulation, 111 2D configurations are identified, all containing a 60{~\AA} vacuum layer. The total energies of the configurations are calculated using the all-electron full-potential FHI-aims code~\cite{blum2009a,huhn2020a,yu2021a} with the Perdew-Burke-Ernzerhof (PBE) exchange-correlation functional~\cite{Perdew1996a}. The tight  numerical settings and a 4$\times$4$\times$1 k-point grids are used in these calculations. All structures are fully optimized using the Broyden-Fletcher-Goldfarb-Shanno (BFGS)~\cite{nocedal2006a} algorithm, with a maximum force component below 10$^{-3}$~eV/{~\AA}.

For their structural, electronic, and vibrational properties, we carried out first-principles calculations based on density functional theory (DFT) using Vienna ab \textit{initio} simulation package (VASP)~\cite{Kresse1996a,Kresse1996b}. Projector augmented wave potentials~\cite{Blochl1994a} were employed to describe the interaction between ions and valence electrons. The generalized gradient approximation (GGA) of PBE~\cite{Perdew1996a} was used as the exchange correlation functional. The cutoff energy for the plane wave basis was chosen to be 400~eV. The Brillouin zone (BZ) was sampled using $\Gamma$-centered $8{\times}8{\times}1$ grid. To avoid the spurious inter-layer interaction, we introduced a vacuum region of 20{~\AA} along the $c$-axis perpendicular to the sheet.  Atomic relaxations were done until the Helmann-Feynman force acting on every atom became smaller than 0.01~eV/{\AA}. SOC is considered for all calculations. The vibrational property of CaP$_3$ was evaluated using the harmonic approximation implemented in PHONOPY package~\cite{Togo2015a}. We used $3{\times}3{\times}$1 supercell structures for the optimized structure of CaP$_3$. The dynamics were studied via canonical ab \textit{initio} molecular dynamics (MD) simulations with 1~fs time steps. To confirm the stability at a higher temperature than room temperature, we used 500~K for 7~ps to reach thermal equilibrium with a $6{\times}6{\times}1$ supercell (128 atoms). 

\section{Results and Discussion}

\begin{figure*}
\centering
\includegraphics[width=1.0\textwidth]{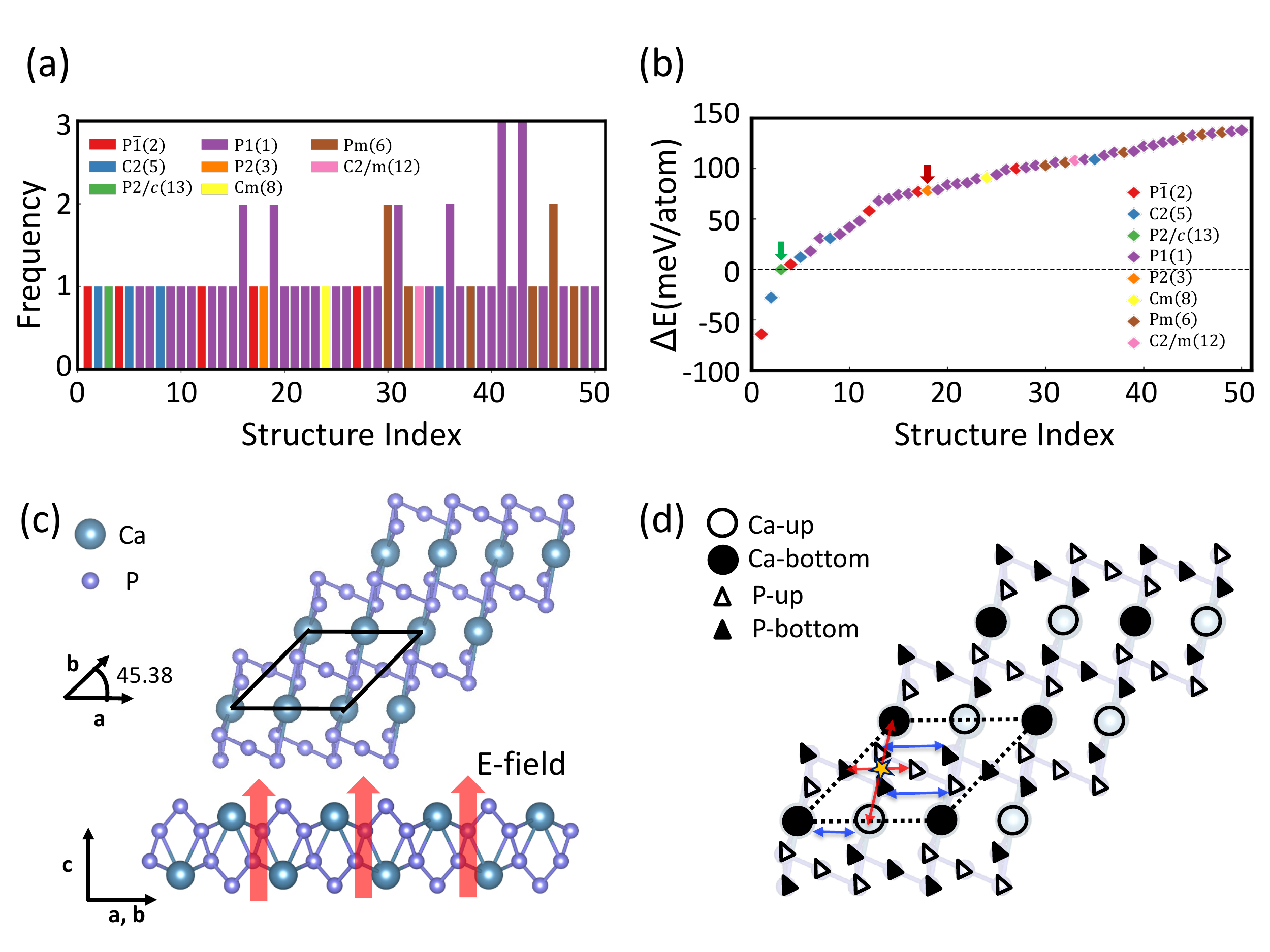}
\caption{
Stable configurations and crystal structures of topological CaP$_3$ ($P2/c$). (a) The low-energy structures of 2D CaP$_3$ found by the PSO algorithm. The crystal structures are color coded and ordered according to the energy hierarchy, i.e., the first one is the most stable structure. The $y$-axis represents the number of occurrences of a structure with the same energy and a space group. The CaP$_3$ with $P2/c$ symmetry is the topological Dirac semimetal identified in this study. (b) The energy ($\Delta$E) of the polymorphs with respect to the $P2/c$ structure (green arrow), with the red arrow indicating the previous reported structure~\cite{ning2018a}. The CaP$_3$ with $P\bar{\text{1}}$ and $C2$ is more stable for $P2/c$, the structure of this study. (c) Top and side views of the $P2/c$ structure representing a distorted hexagonal lattice (highlighted by the unit cell of the black line). (d) The circles and triangles represent Ca and P atoms, respectively, with the solids and hollows representing below and above the mirror plane ($M_z$) on the $xy$-plane, respectively. The orange star is the inversion point, and the red and blue arrows denote atoms corresponding to the inversion and asymmetric symmetry of certain atoms in unit cell. 
\label{Fig1}}
\end{figure*}

The evolutionary algorithm based on the first-principles DFT calculations identified various stable crystal structures of CaP$_3$. Among them we compiled the first 50 most stable structures in the structural frequencies in Fig.~\ref{Fig1}(a) and the energy distributions of the polymorphs in Fig.~\ref{Fig1}(b). The most stable phase found in our study is a $\sim$1~eV bandgap semiconductor with $P\bar{\text{1}}$ symmetry, and the second most stable phase is also a $\sim$1~eV band gap semiconductor with $C2$ symmetry with energy higher by 35~meV/atom (See their atomic structures and band structure in SI-Fig. 1. We confirm that these two lowest structures have no interesting topological features. The next stable structure has $P2/c$ symmetry with an energy 25~meV/atom higher than the $C2$ structure, but it is even more stable than the previous reported $P\bar{\text{1}}(2)$ phase with 75~meV/atom~\cite{ning2018a}. Among these phases, we found the $P2/c$ phase to be the most interesting - the crystal symmetry exhibits both inversion and nonsymorphic symmetries. From now on, we focus on the $P2/c$ phase and refer to it as the CaP$_3$ unless otherwise stated. The CaP$_3$ consists of a distorted hexagonal unit cell with an angle of $\sim$45$^\circ$ between the cell vectors as shown in Fig.~\ref{Fig1}(c). Figure~\ref{Fig1}(d) analyzes the crystal symmetries, where the orange star is an inversion center for the symmetry operation (indicated by the red arrows) and the blue arrows indicate a nonsymmorphic symmetry consisting of mirror operation of the $xy$-plane with half translation along the $x$-axis. The optimized atomic positions are listed in SI-Fig. 2(a).

\begin{figure*}
\centering
\includegraphics[width=1.0\textwidth]{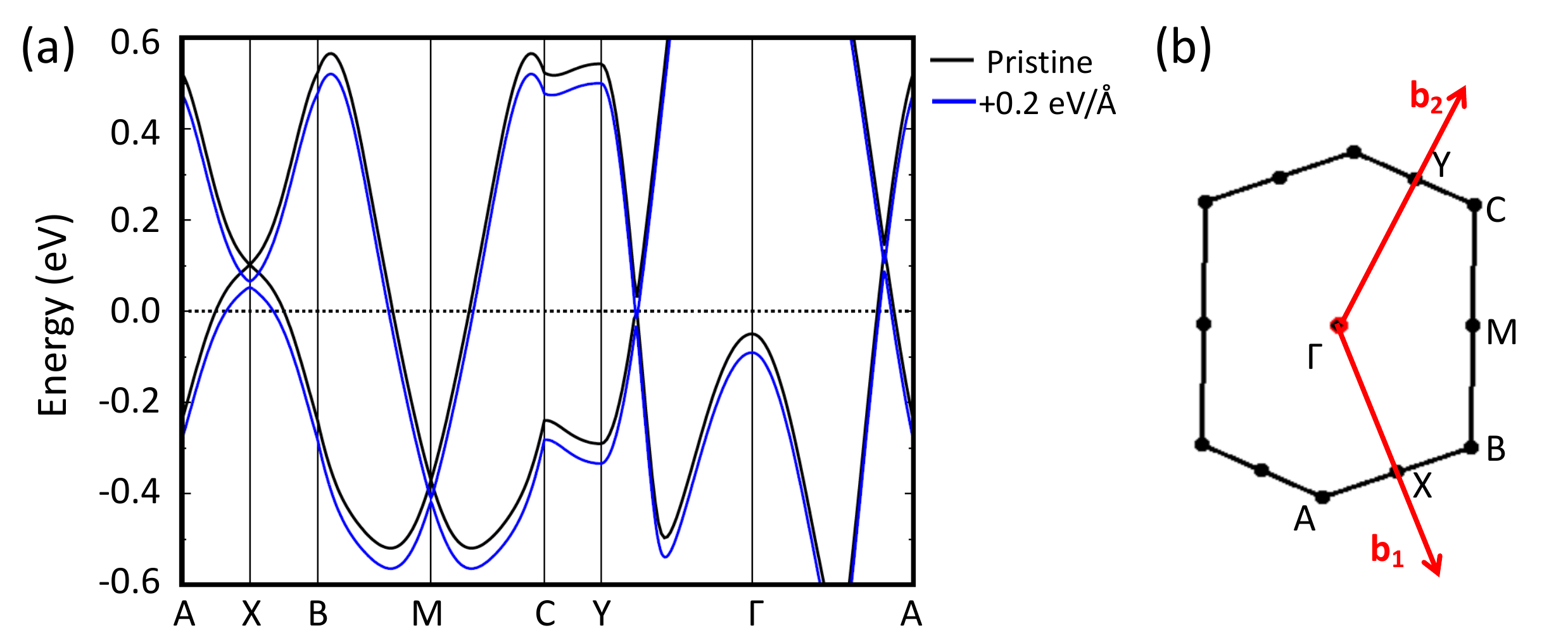}
\caption{
Electronic energy bands of the pristine CaP$_3$ (black line) compared to those with +0.2~eV/{~\AA} electric field applied perpendicular to the surface (blue line). The Brillouin zone shows the band paths along the high-symmetry points. 
\label{Fig2}}
\end{figure*}

Next, we investigate the electronic properties of CaP$_3$. Figure~\ref{Fig2}(a) shows the electronic band structures of CaP$_3$ with SOC along the high-symmetry line in BZ as shown in Fig.~\ref{Fig2}(b). The CaP$_3$ exhibits metallic state with two bands crossing at the X- and M-points near the Fermi level. The presence of both inversion ($P$) and time-reversal ($\Theta$) symmetries guarantees the two-fold degenerate band with SOC at all BZ. Therefore, the crossing points at the X- and M-points are the locations where the four-fold degenerated Dirac state exists. The band structure also represents a small SOC effect for light elements and yields essentially the same eigenvalues regardless of SOC. 

We then perform a symmetry analysis to explain the protection of the Dirac points at the X- and M-points. In the following analysis, we include the electron spin and the SOC. The crystal symmetries underlying the CaP$_3$, characterized by the inversion ($P$) and nonsymmorphic operations ($\widetilde{M}_{z}$), can be decomposed into products of point and translation group operations as $\widetilde{M}_{z}=M_{z}T_{a\hat{x}/2}$~\cite{young2015a}. Here, $M_{z}$ is the mirror operation about the $xy$-plane and $T_{a\hat{x}/2}$ is the translation by half translation along the $x$-axis, as shown in Fig.~\ref{Fig1}(d). For the nonsymmorphic symmetry, the point group operator should commute with the translation group operation. The representation of $T_{a\hat{x}/2}$ for a Bloch state with $k_{x}$ is $T_{a\hat{x}/2}=e^{ik_xa/2}$, that is, $\widetilde{M}_{z}$ becomes $\widetilde{M}_{z}=M_{z}e^{ik_xa/2}$. To find the eigenvalues of $\widetilde{M}_{z}$, we note that 
\[
  \left({\widetilde{M}_{z}}\right)^{2}={M_{z}}^{2}e^{ik_xa}=-e^{ik_xa},
\]
where $e^{ik_xa}$ denotes the translation by one unit cell along $x$ directions and  ${M_z}^{2}$ is -1 from the 2$\pi$ rotation on spin. The spinor representation of $M_z$ is  given by $M_z=i\sigma_z{\otimes}R_z(\pi)P$, where $\sigma_z$ is the $z$ component of the Pauli spin matrices affecting only the spin parts of the wavefunction ; $R_z(\pi)$ and $P$ are, respectively, a real space rotation around the $z$-axis by an angle $\pi$ and an inversion $\mathbf{r}\rightarrow-\mathbf{r}$ affecting only the orbital parts of the wavefunction. We use that the mirror operation is a multiple of two operations, an inversion followed by a $\pi$ rotation, and that the representation of an inversion in the spin-half space is the identity. The eigenvalues resulting from the solution of the, $\widetilde{M}_{z}\psi=E_{z}\psi$ are $E_{z}=\pm{i}e^{ik_{x}a/2}$. Now, we consider the inversion symmetry ($P$) operation with the nonsymmorphic ($\widetilde{M}_z$) symmetry. These operations, represented as $\widetilde{M}_{z}:(x,y,z)\rightarrow(x+1/2,y,-z)$ and $P:(x,y,z)\rightarrow(-x,-y,-z)$, do not commute each other. For the time-reversal symmetry operation with $\widetilde{M}_{z}P$ and $P\widetilde{M}_{z}$ holds $\widetilde{M}_{z}P\Theta:(x,y,z)\rightarrow(-x+1/2,-y,z)$ and $P\Theta\widetilde{M}_{z}:(x,y,z)\rightarrow(-x-1/2,-y,z)$, respectively. Therefore, $\widetilde{M}_{z}P=e^{ik_{x}a}P\widetilde{M}_{z}$ and $\widetilde{M}_{z}P\Theta=e^{ik_{x}a} P\Theta\widetilde{M}_{z}$ and, for an eigenstate $\ket{E_{z}}$ of with eigenvalue $E_z$, it becomes $\widetilde{M}_{z}(P\Theta\ket{E_{z}})=e^{ik_{x}a}\pm{i}e^{ik_{x}a/2}(P\Theta\ket{E_{z}})=-E_{z}(P\Theta\ket{E_{z}})$. Due to $P$ and $\Theta$ symmetries under SOC, all bands are two-fold degenerate with $\ket{E_z}$ and $P\Theta\ket{E_z}$ at the whole k-points. For $k_x=\pi/a$, $E_z$ is $\pm{1}$ and the $\left[\widetilde{M}_{z}, P\right]$ is $0$. Two space group (symmetric) operators that anticommute guarantee two-fold degeneracy~\cite{wieder2016a}. Moreover, X- and M-points are TRIM points, therefore all states $\ket{Ez}$ at the X-point have another degenerate Kramer's pair $\Theta\ket{E_{z}}$ with the same eigenvalue $E_{z}$. The four eigenstates such as $\ket{E_z}$, $\Theta\ket{E_z}$, $P\Theta\ket{E_z}$, and $P\ket{E_z}$ , are degenerate at X- and M-points. Except for the TRIM points, X-, Y-, and M-points, it is not invariant with the $\Theta$ operation, so the four-fold degeneracy is broken by two doubly degenerate bands. For the Y-point, $k_x$=0 and $k_{y}=\pi/a$, the anticommute for $\widetilde{M}_z$ and $P$ is not satisfied, so it has only a two-fold degeneracy for the inversion symmetry.

Our symmetry analysis predicts the four-fold degenerate Dirac states at the X- and M-points. To verify the theoretical prediction, we perform first-principles calculations of the band structures. In particular, we break the inversion symmetry of the CaP$_3$ by applying an electric field perpendicular to the surface in Fig.~\ref{Fig1}(c). In principle, any field strength breaks the symmetry, leading to the splitting of degenerate states, if present. In practice, however, the splitting is difficult to detect if the field strength is too low. Therefore, we choose a large field strength, +0.2~eV/{\AA}, as an example. Figure~\ref{Fig2}(a) compares the changes in the band structures due to the electric field. The E-field indeed breaks the inversion symmetry, leading to the splitting of two doubly degenerate bands in the X- and M-points, which are protected by the time-reversal symmetry. These results also show the band structures of CaP$_3$ with full BZ in SI-Fig. 2(b) and the other alkaline earth metal trinictogenide such as SrP$_3$ and BaP$_3$ as shown in SI-Fig. 3. 

\begin{figure*}
\centering
\includegraphics[width=1.0\textwidth]{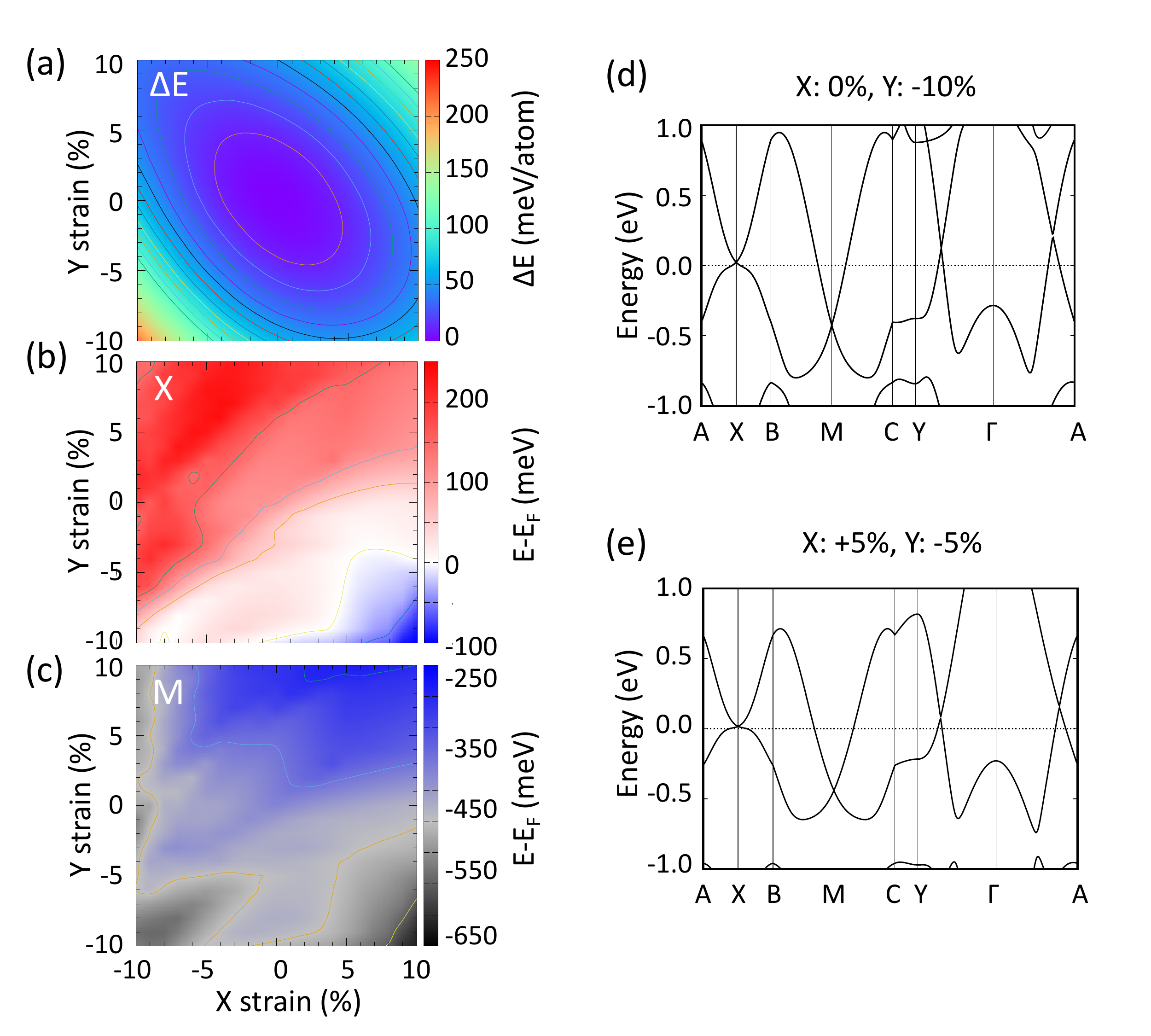}
\caption{
The effects of strains on the stability and electronic properties of CaP$_3$. (a) Changes in the total energy (eV/atom) of the structure by applying strains, where the energy is given with respect to the equilibrium structure. (b), (c) The energy level of the Dirac state with respect to the Fermi level as a function of the strains at X- and M-points, respectively. (d), (e) Band structures with uniaxial compressive strain (X:0\%, Y:-10\%) and biaxial strain (X:+5\%, Y:-5\%), respectively.
\label{Fig3}}
\end{figure*}

The number of electrons of CaP$_3$ satisfies the electron filling condition (4$n$+2, $n$=11)~\cite{young2015a,wieder2016a} , leading to the Dirac states near the Fermi level$-$ +100~meV and -400~meV at X- and M-points, respectively. To move the Dirac state closer to the Fermi level, some external perturbations are required, such as the application of an E-field. However, the electric field is not ideal since it breaks the spatial inversion symmetry. Therefore, we focus on the strain to preserve the crystal symmetry as shown in Fig.~\ref{Fig3}. Since the Dirac state of CaP$_3$ is guaranteed by the inversion symmetry, the time-reversal symmetry, and the nonsymmorphic symmetry, the Dirac state cannot be broken if these symmetries are preserved. In Fig.~\ref{Fig3}(a), we first estimate the energy costs for different strains. The application of strains up to about +5\% increases the energy by 30~meV, which corresponds to a thermal energy variation of ~300~K. Therefore, we assume that these structures are stable at room temperature. The energy levels of the Dirac states at the X- and M-points can be tuned by stretching as shown in Figs.~\ref{Fig3}(b) and (c)). We find that the energy level of the Dirac state of the X-point is mobile and can find the Fermi level by strain, whereas the energy level of the M-point remains more than 250~meV below the Fermi level regardless of strain. The Dirac state of the X-point is pinned to the E$_{F}$ at certain strains, such as 10\% compressive strain in the $y$ direction and 5\% tensile strain in the $x$ direction with 5\% compressive strain in the $y$ direction in Fig.~\ref{Fig3}(b). In Figs.~\ref{Fig3}(d) and (e), it is confirmed that the Dirac state is pinned to the E$_{F}$; the symmetries protecting the Dirac state are robust against various strains, such as biaxial, uniaxial, and shear deformation in the plane.  

\begin{figure*}
\centering
\includegraphics[width=1.0\textwidth]{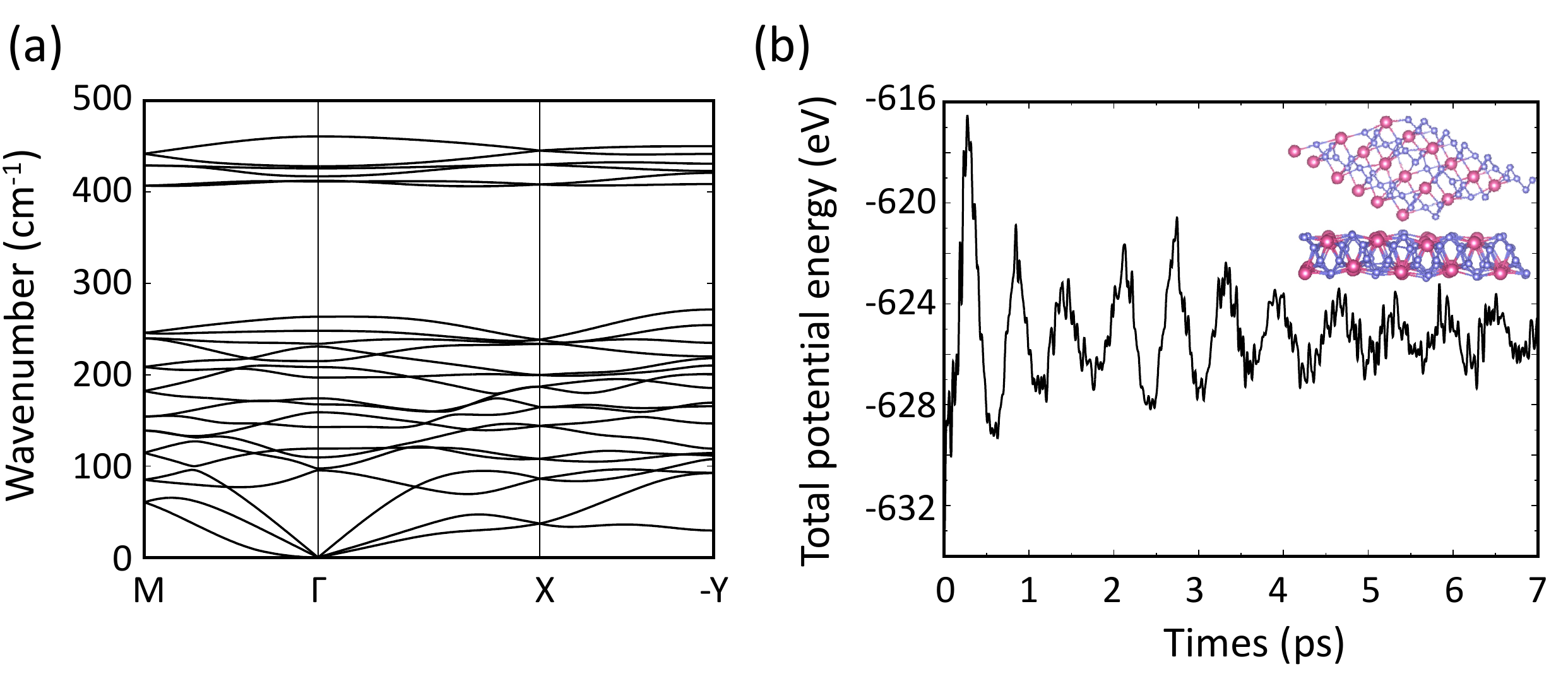}
\caption{
The stability of CaP$_3$. (a) Phonon band structures along the high-symmetry points shown in Fig.~\ref{Fig2}. (b) Total potential energy as a function of time during canonical MD simulations at 500~K with the final structural at the end of the simulation time of 7~ps (inset). 
\label{Fig4}}
\end{figure*}

Regardless of the cohesive energy, a structure can be considered stable only if it does not change spontaneously. The spontaneous structural change can be due to a negative frequency in the vibrational band dispersion. To verify the stability and structural rigidity for CaP$_3$, we investigate the vibration spectra of its structure as presented in the Fig.~\ref{Fig4}(a). The absence of a negative frequency in the phonon band means that CaP$_3$ is dynamically stable. Since the slope of the longitudinal acoustic (LA) branch near the $\Gamma$ is isotropic, the rigidity in the $\Gamma-$M direction and the $\Gamma-$X direction is similar. While phonon dispersion can predict structural stability in the harmonic region, it cannot determine whether or not the structure collapses at a finite temperature. To evaluate the thermodynamic stability of the new CaP$_3$ structure, we perform canonical molecular dynamics (MD) simulations at a high temperature, 500~K. Figure~\ref{Fig4}(b) presents the fluctuations of the total potential energy for the 7~ps MD simulations and the final snapshot of the structure. Our results show that the CaP$_3$ structure fluctuates at 500~K, but it is thermodynamically stable, and the Dirac state is even robust at this high temperature.

\section{Conclusion}
We use an evolution algorithm in combination with first-principles density functional theory calculations and find a new Dirac semimetal 2D CaP$_3$, composed of light elements. The most intriguing feature of the new phase is that a non-symmorphic and inversion symmetry guarantees symmetry-protected degeneracies in the electronic band structures, including a fourfold degenerate Dirac state that is intact under small strains. The Dirac state near the Fermi level as a result of satisfying the 4$n$+2 electron filling criterion with the location of the Dirac state controlled by strains. The kinetic and thermodynamic stability of the new phase is confirmed by phonon dispersion analysis and MD simulations. 

\section*{Acknowledgement} 
Theory work is supported by the US Department of Energy, Office of Science, Office of Basic Energy Sciences, Materials Sciences and Engineering Division (W. L., S. Y., Y. Z., and M. Y.), and by the U.S. Department of Energy (DOE), Office of Science, National Quantum Information Science Research Centers, Quantum Science Center (S.-H. K.). This research used resources of the Oak Ridge Leadership Computing Facility and the National Energy Research Scientific Computing Center, US Department of Energy Office of Science User Facilities.

This manuscript has been authored by UT-Battelle, LLC, under Contract No. DE-AC0500OR22725 with the U.S. Department of Energy. The United States Government retains and the publisher, by accepting the article for publication, acknowledges that the United States Government retains a non-exclusive, paid-up, irrevocable, world-wide license to publish or reproduce the published form of this manuscript, or allow others to do so, for the United States Government purposes. The Department of Energy will provide public access to these results of federally sponsored research in accordance with the DOE Public Access Plan (http://energy.gov/downloads/doe-public-access-plan).

\clearpage
\bibliography{biblio}

\clearpage
\section*{Supplemental information} 
\renewcommand{\thefigure}{S\arabic{figure}}
\setcounter{figure}{0}

\begin{figure*}
\centering
\includegraphics[width=1.0\textwidth]{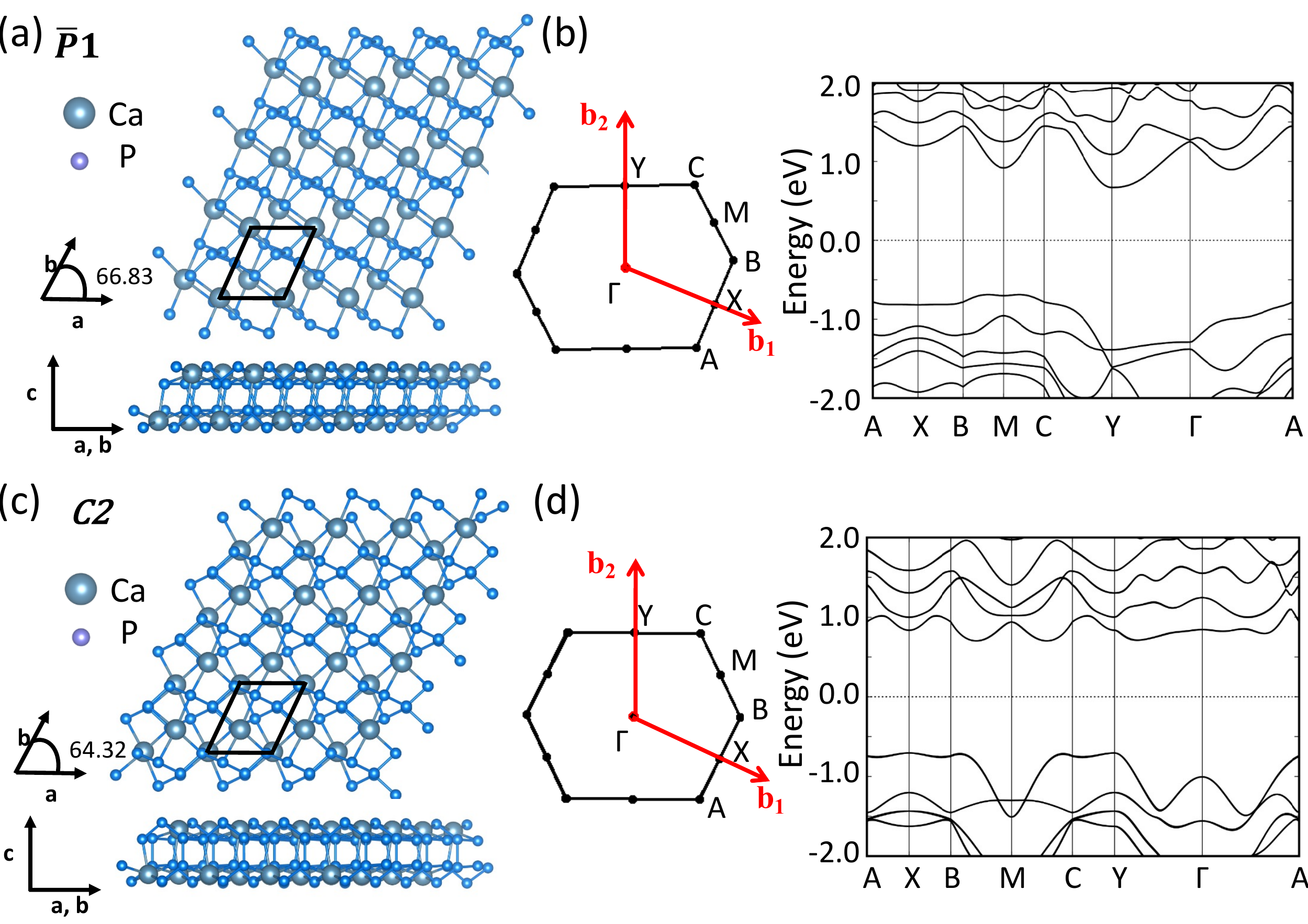}
\caption{
Atomic structures of the new CaP$_3$ with $P\bar{\text{1}}$(a) and $C2$(c) symmetries and their electronic bands, (b) and (d), with high symmetry points in the respective Brillouin zones.   
\label{SI-Fig1}}
\end{figure*}

\begin{figure*}
\centering
\includegraphics[width=1.0\textwidth]{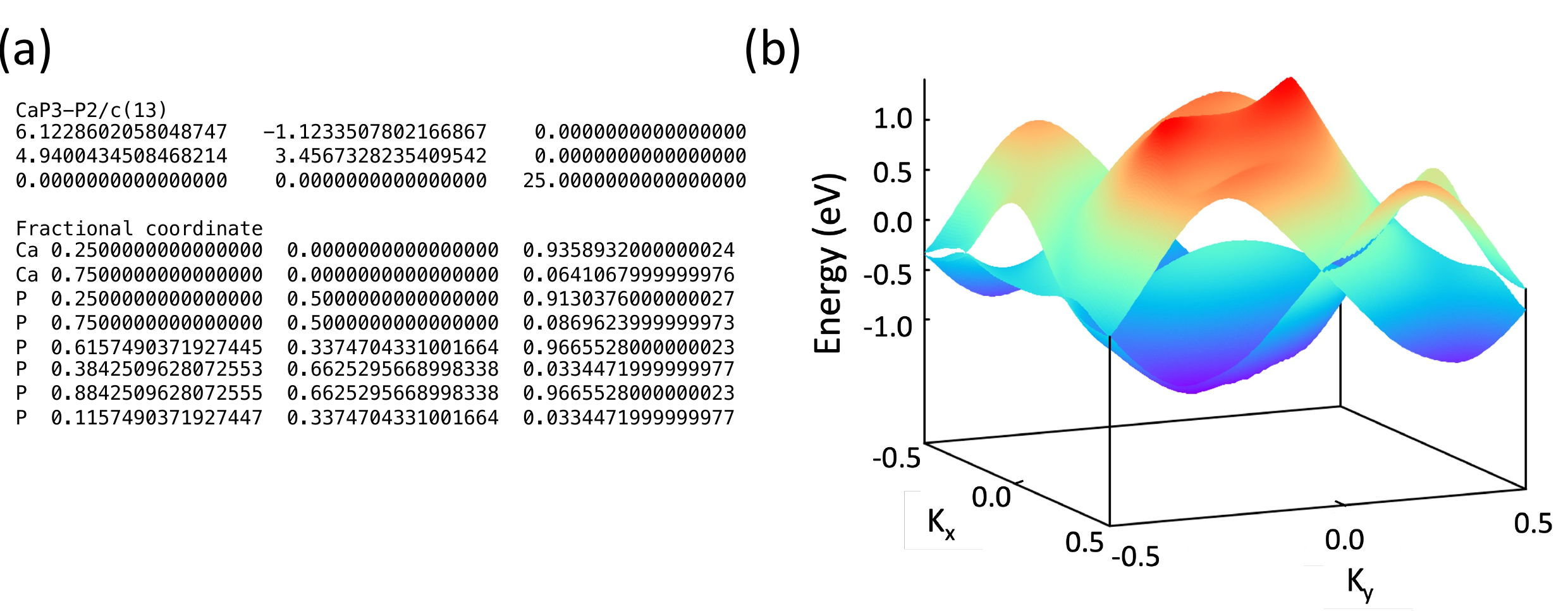}
\caption{
Atomic coordinates (a) and band structure (b) of topological CaP$_3$ with $P2/c$ symmetry.
\label{SI-Fig2}}
\end{figure*}

\begin{figure*}
\centering
\includegraphics[width=1.0\textwidth]{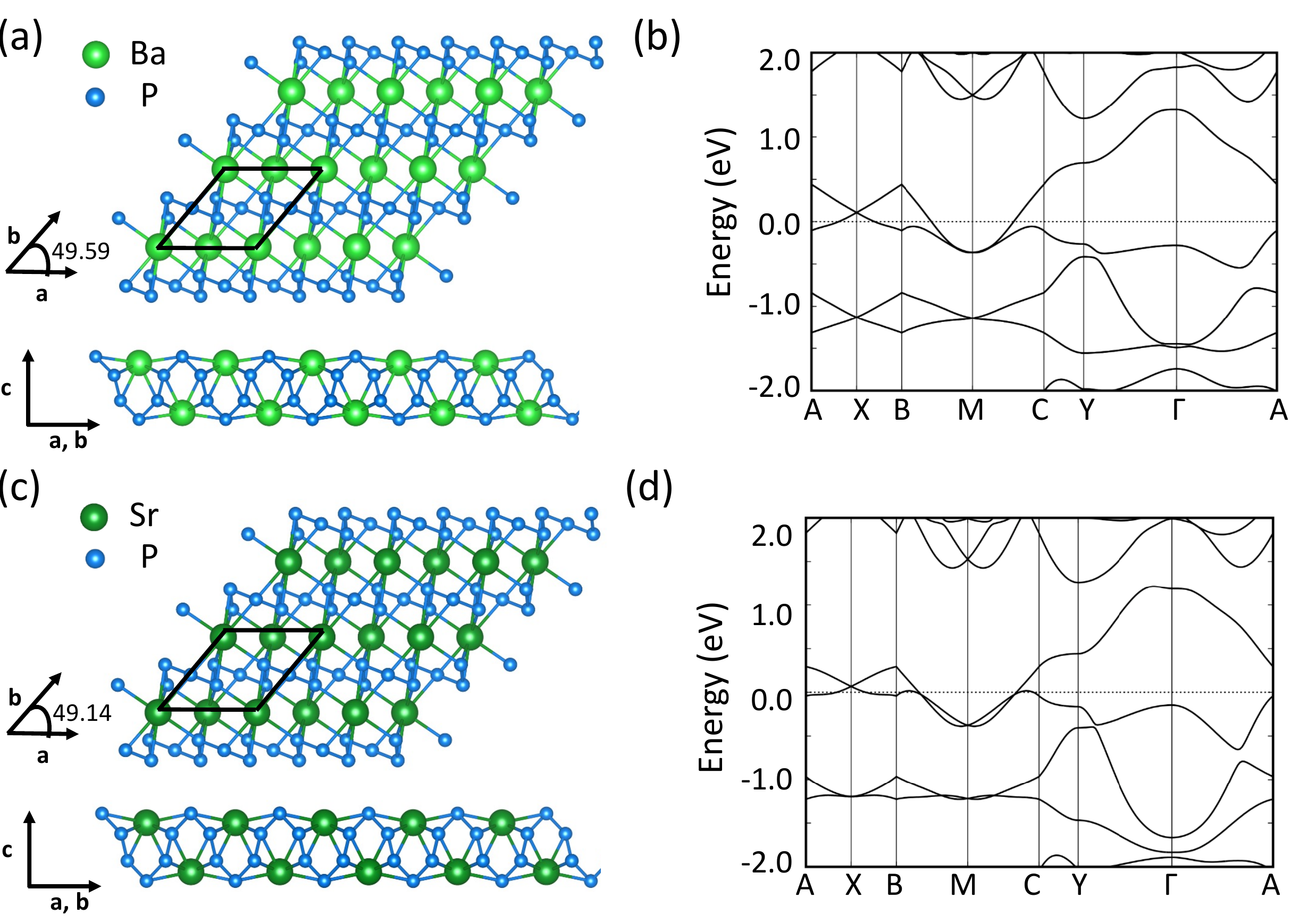}
\caption{
Atomic and band structures of BaP$_3$, (a) and (b), SrP$_3$ (c) and (d).
\label{SI-Fig3}}
\end{figure*}

\end{document}